\documentclass{sty/IEEEtran}

\usepackage{amsmath}
\usepackage{comment}
\usepackage{amsfonts}
\usepackage{amssymb}
\usepackage{graphicx}
\usepackage{algorithm}
\usepackage{algpseudocode}
\ifCLASSOPTIONcompsoc
\usepackage[caption=false, font=normalsize, labelfont=sf, textfont=sf]{subfig}
\else
\usepackage[caption=false, font=footnotesize]{subfig}
\fi

\usepackage{cite}

\usepackage{threeparttable}
\usepackage{multirow}
\usepackage{times}
\usepackage{float}
\usepackage{enumerate}
\usepackage{hyperref}
\usepackage[none]{hyphenat}
\usepackage[normalem]{ulem}
\usepackage{color}
\usepackage{xcolor}
\usepackage{url}
\usepackage{tikz}
\usepackage{pgfplots}
\pgfplotsset{compat=newest}
\usetikzlibrary{positioning}

\newcommand{\qed}{\nobreak \ifvmode \relax \else
	\ifdim\lastskip<1.5em \hskip-\lastskip
	\hskip22em plus0em minus0.5em \fi \nobreak
	\vrule height0.4em width0.3em depth0.25em\fi}

\newlength\fheight
\newlength\fwidth
\begin{document}

    \title{
        {\fontsize{24}{26}\selectfont{Communication\rule{29.9pc}{0.5pt}}}\break\fontsize{16}{18}\selectfont
        Fast Plasma Frequency Sweep in Drude-like EM~Scatterers via the Reduced-Basis Method
    }
	\author{Clara Iglesias-Tesouro, Valent\'{i}n de la Rubia, Alessio Monti and Filiberto Bilotti
    \thanks{C. Iglesias-Tesouro and V. de la Rubia are with the Departamento de Matem\'{a}tica Aplicada a las {TIC}, ETSI de Telecomunicaci\'{o}n, Universidad Polit\'{e}cnica de Madrid, 28040 Madrid, Spain (e-mails: clara.iglesias.tesouro@upm.es; valentin.delarubia@upm.es).}
    \thanks{A. Monti is with the Department of Industrial, Electronic and Mechanical Engineering, Roma Tre University, 00146 Rome, Italy (e-mail: alessio.monti@uniroma3.it).}
    \thanks{F. Bilotti is with the Department of Industrial, Electronic and Mechanical Engineering, Roma Tre University, 00146 Rome, Italy and also with the National Inter-University Consortium for Telecommunications (CNIT), 43124 Parma, Italy (e-mail: filiberto.bilotti@uniroma3.it).}    
	}
	\maketitle
	\begin{abstract}
        In this work, we propose to use the Reduced-Basis Method (RBM) as a model order reduction approach to solve Maxwell's equations in electromagnetic (EM) scatterers based on plasma to build a metasurface, taking into account a parameter, namely, the plasma frequency. We build up the reduced-order model in an adaptive fashion following a greedy algorithm. This method enables a fast sweep over a wide range of plasma frequencies, thus providing an efficient way to characterize electromagnetic structures based on Drude-like plasma scatterers. We validate and test the proposed technique on several plasma metasurfaces and compare it with the finite element method (FEM) approach.

	\end{abstract}
	\markboth{}{Iglesias-Tesouro and de la Rubia: Fast Plasma Frequency Sweep in Drude-like Scatterers via the Reduced-Basis Method in Electromagnetics}
	\IEEEpeerreviewmaketitle
\section{Introduction}
\label{Sec-Introduction}
%
Conventional high-fidelity methods (HFMs), such as the finite element method (FEM), have proven to be a very accurate approach for the numerical resolution of PDE-governed physical problems. Their applicability to domains with complex geometries and its high reliability have made it possible to solve many non-analytical problems in the fields of physics and engineering with great precision {\color{black}\cite{Monk,Jin}}. However, the design and optimization of products in engineering applications often requires modifications and successive simulations, which, given the high computational costs of HFMs, is sometimes not feasible.

Instead, model order reduction (MOR) techniques have turned out to be a particularly useful alternative \cite{benner2015survey}. Their power to compress, with good approximation, high-dimensional problems to their reduced equivalents has constituted a milestone in computational analysis and numerical simulations. In particular, \textcolor{black}{the Reduced-Basis Method (RBM) \cite{Noor1979ReducedBT, patera2001}} is a common choice as a MOR approach. Given its efficiency and reliability, it had become a widely used technique in areas such as \textcolor{black}{fluid mechanics, thermodynamics and electromagnetics \cite{QuarteroniManzoniRBM, HesthavenRozzaRBM}.}

In this work, we apply RBM in finite element approximations to Maxwell's equations, in order to obtain the electromagnetic response of different Drude-like EM scatterers. Such structures have come up as very promising elements (\textit{meta-atoms}) in the design of Huygens metasurfaces (HMSs) \cite{HMSsChen, HMSsEpstein}. \textcolor{black}{This special type of metasurfaces are 2D-arrays of metamaterials which, based on the Huygens principle, can provide full-wave transmission and almost total phase coverage at a wide range of frequencies \cite{PfeifferCascadedMW, MontiCascadedMW, DlinDielectricOP, ChongDielectricOP}. Given recent technological challenges, not only traditional metasurfaces with fixed functionalities have been developed, but also reconfigurable metasurfaces are being explored \cite{RamacciaReconfigurabilityStateOfArt}. In this way, by tuning different characteristics of the metasurfaces \cite{SievenpipeTunableSurf, WangTunableMetasurf}, their performance can be adapted to different operational requirements, enabling a dynamic manipulation of electromagnetic waves. 
In recent works \cite{SakaiPlasmas, YuanSlabPlasma}, it has been shown that plasma structures can be a promising metamaterial to control propagation of electromagnetic waves and thus conform novel reconfigurable metasurfaces. For instance, scatterers made with plasma, based on the Drude model, have been used to design reconfigurable phase-gradient metasurfaces for beam steering \cite{monti2024drudelikescatterers}, tunable metasurfaces for frequency band-gap control \cite{SakaiBandGaps}, and tunable plasma photonic crystals \cite{WangPlasmaPhotonic}.} A proper design of such structures requires an adequate characterization of the scatterer EM response as a function of its plasma properties. This characterization can be done with ease by performing a fast plasma frequency sweep via RBM. \textcolor{black}{This approach} relies on the fact that each EM response, over a wide range of plasma frequencies, can be represented by a set of few FEM solutions corresponding to certain plasma frequencies. These EM field solutions constitute the reduced basis for our system, which allows us to turn the high-dimensional FEM space into a low-dimensional approximation space, providing a reduced-order model (ROM) for Maxwell's equations. Thus, the proposed RBM will only require to perform the finite element approximation of the EM field at certain plasma frequencies to obtain the EM response over the entire range of interest, dramatically reducing the computational costs. This approach has been successfully applied in electromagnetics earlier in different scenarios \cite{delarubia2009reliable,delaRubia2018CRBM,morHesB13,morFenB19,Silveira2014Reduced-OrderModels,Edlinger2017finite,EdlingerDrudeModel2017,vidal2018computing,nicolini2019model,Rewienski2016greedy,codecasa2019exploiting,xue2020rapid,szypulski2020SSMMM,chellappa2023infsup,chellappa2021adaptive,fotyga2022MTT,SebastianSchops2023,SebastianSchops2024}.

This work is structured as follows. In Section \ref{Sec-FEM}, we use FEM as an approach to approximate the EM field in our analysis domain and numerically solve Maxwell's equations in plasmas by means of a Drude model. Section \ref{Sec-RBM} presents our RBM approach. In Section \ref{Sec-Examples}, we test the proposed RBM methodology on some Drude-like plasma scatterers. Finally, in Section \ref{Sec-Conclusions}, we comment on the conclusions.
\section{Finite element method for Maxwell's equations}
\label{Sec-FEM}
The weak form of time-harmonic Maxwell's equations is given by \cite{Monk}
\begin{equation}
    \begin{split}
        & \text{Find}\quad \vec{E}\in \mathcal{H}\quad \text{ such that:} \\
        & a(\vec{E}, \vec{\nu})=f(\vec{\nu})\quad \forall\vec{\nu}\in\mathcal{H},
    \end{split}
    \label{ec: weak_form}
\end{equation}
where $\mathcal{H}$ is an admissible Hilbert space and the bilinear and linear forms read
\begin{equation}
    \begin{aligned}
    a(\vec{E},\vec{\nu})&=\int_{\Omega}\frac{1}{\mu}(\nabla\times\vec{E})\cdot(\nabla\times\vec{\nu})\ dx-\omega^2\int_{\Omega}\varepsilon\vec{E}\cdot\vec{\nu}\ dx\\
    \label{ec: a}
    f(\vec{\nu})&= j\omega\int_{\partial\Omega}\vec{J}\cdot\vec{\nu}\ ds,
    \end{aligned}
\end{equation}
where $\vec{E}$ is the electric field, $\vec{J}$ is the excitation current, $\Omega \subset \mathbb{R}^3$ is the analysis domain, $\varepsilon$ and $\mu$ are the dielectric permittivity and magnetic permeability, respectively, and $\omega$ is related to the operating frequency f ($\omega=2\pi$f). As a slight abuse of notation, from now on we will refer to both $\omega$ and f as frequency. The finite element method for solving this boundary-valued electromagnetic problem is based on approximating the infinite-dimensional Hilbert space $\mathcal{H}$, where electric field functions reside, by a finite-dimensional space, $V_h$, given by N\'{e}d\'{e}lec finite element functions for the electric field (1-forms), $\{\vec{f}_{k}, k = 1, \dots, n\}$, with $n$ the dimension of the finite element space $V_h$ ($n = \operatorname{dim}(V_h)$), i.e., number of degrees of freedom in the FEM approximation. Therefore, the electric field can be written by an approximation in the finite element basis,
\begin{equation}
  \vec{E}\simeq \sum_{k=1}^{n}x_k\vec{f}_{k}.
  \label{ec: E}
\end{equation}
With this approach, the variational problem \eqref{ec: weak_form} turns into a system of linear equations, whose solution vector $\mathbf{x}$ contains the set of $x_k$ coefficients to describe the electric field in the finite element basis, thus,
\begin{equation}
    \left(\mathcal{S}+s^2\mathcal{T}\right)\mathbf{x} = \mathbf{b}.
    \label{ec: FEM}
\end{equation}
The \textit{stiffness} and \textit{mass} matrices, $\mathcal{S}$ and $\mathcal{T}$, are given below in equations \eqref{ec: S} and \eqref{ec: T}, respectively. The \textit{excitation} vector, $\mathbf{b}$, is detailed in \eqref{ec: b} and $s= j\omega$.
\begin{equation}
    \mathcal{S}_{ij}\equiv\int_{\Omega}\frac{1}{\mu}\left(\nabla\times\vec{f}_{i}\right)\cdot \left(\nabla\times\vec{f}_{j}\right)dx.
    \label{ec: S}
\end{equation}
\begin{equation}
    \mathcal{T}_{ij} \equiv\int_{\Omega}\varepsilon\vec{f}_{i}\cdot\vec{f}_{j}dx.
    \label{ec: T}
\end{equation}
\begin{equation}
    b_i=j\omega\int_{\Omega}\vec{J}\cdot\vec{f}_{i}dx.
    \label{ec: b}
\end{equation}
For an analysis domain $\Omega$ consisting of different dielectric materials, each of which characterized by its dielectric permittivity, $\varepsilon$, the \textit{mass} matrix $\mathcal{T}$ in \eqref{ec: T} can be broken down into different subdomain \textit{mass} matrices, for instance, $\mathcal{T}=\varepsilon_1\mathcal{T}_1+\varepsilon_2\mathcal{T}_2+\varepsilon_3\mathcal{T}_3$, where three different decomposition subdomains have been taken into account. For the sake of discussion, let us consider the specific case of a plasma subdomain $\Omega_p \subset \Omega$, with plasma permittivity $\varepsilon_p(\omega_p)$, given by the Drude model \eqref{ec: epsilon_p}, cf. \cite{Drude}. Assuming this plasma subdomain is embedded in another material in the remaining analysis domain $\Omega \setminus \Omega_p$, with dielectric permittivity $\varepsilon_m$, the matrix problem \eqref{ec: FEM} can be written down as follows,
\begin{equation}
    \left(\mathcal{S}+s^2(\varepsilon_m\mathcal{T}_m+\varepsilon_p(\omega_p)\mathcal{T}_p)\right)\mathbf{x}=\mathbf{b},
    \label{ec: FEM_2}
\end{equation}
where
\begin{equation}
    \label{ec: epsilon_p}
    \frac{\varepsilon_p(\omega_p)}{\varepsilon_0} = 1-\frac{\omega_p^2}{\omega^2-j\omega\gamma(\omega_p)}.
\end{equation}
$\varepsilon_0$ is the dielectric permittivity in vacuum and $\gamma$ is the collision frequency, $\gamma(\omega_p)=\kappa\omega_p$, proportional to the plasma frequency, $\omega_p$. 
Note that the original \textit{mass} matrix $\mathcal{T}$ is broken down into $\mathcal{T}=\varepsilon_m\mathcal{T}_m+\varepsilon_p\mathcal{T}_p$ with
\begin{equation}
    \mathcal{T}_{p_{ij}} \equiv\int_{\Omega_p}\vec{f}_{i}\cdot\vec{f}_{j}dx
    \label{ec: T_p}
\end{equation}
\begin{equation}
    \mathcal{T}_{m_{ij}} \equiv\int_{\Omega \setminus \Omega_p}\vec{f}_{i}\cdot\vec{f}_{j}dx.
    \label{ec: T_m}
\end{equation}

By defining $\mathbf{S}\equiv \mathcal{S}+s^2\varepsilon_m\mathcal{T}_m$ and $\mathbf{T}\equiv s^2\mathcal{T}_p$, we have a new system of equations that, for a fixed electromagnetic frequency $\omega$ and non-plasma material, $\varepsilon_m$, can be parametrized using only the plasma frequency $\omega_p$ (which changes the plasma permittivity $\varepsilon_p$), thus,
\begin{equation}
    \left(\mathbf{S}+\varepsilon_p(\omega_p)\mathbf{T}\right)\mathbf{x}=\mathbf{b}.
    \label{ec: FEM_plasma}
\end{equation}
As a result, we can obtain the electric field at any plasma frequency $\omega_p$ in our band of interest, $\mathcal{B}$, by solving the vector $\mathbf{x}$ from equation \eqref{ec: FEM_plasma}, provided that we build up the matrices $\mathbf{S}$, $\mathbf{T}$ and $\mathbf{b}$ that determine the physics of the device under consideration. In general, FEM problems in electromagnetics require very large approximation dimensions, involving matrices in the order of $10^6$. Hence, solving \eqref{ec: FEM_plasma} over a wide range of plasma frequencies may require high computational costs. As an alternative, we propose an RBM procedure as MOR approach which allows us to obtain the EM field solutions in a broad band of plasma frequency values in a fast and reliable way.

\section{Reduced-Basis Method}
\label{Sec-RBM}
RBM consists on approximating the large-dimensional FEM space $V_h$ by a reduced space of dimension $N$ ($N\ll n = \operatorname{dim}(V_h)$), detailed by the reduced basis $\{\mathbf{v}_i, i = 1,\dots,N\}$, which defines the reduced-basis projection matrix $\mathbf{V}=\operatorname{cols}\lbrace \mathbf{v}_i\rbrace_{i=1}^{N}$. As a result, we can approximate the solution vector $\mathbf{x}$ in \eqref{ec: FEM_plasma} as
\begin{equation}
    \mathbf{x}(\omega_p)\simeq \mathbf{V}\mathbf{\tilde{x}}(\omega_p)=\sum_i^N{\tilde{x}}_i(\omega_p)\mathbf{v}_i.
    \label{ec: x-RBM}
\end{equation}
By using the approximation \eqref{ec: x-RBM} in \eqref{ec: FEM_plasma}, and multiplying \eqref{ec: FEM_plasma} by $\mathbf{V}^T$, we obtain an analogous reduced system of equations of dimension $N  \ll n$, whose solution vector $\mathbf{\tilde{x}}(\omega_p)$ contains the $\tilde{x}_i(\omega_p)$ coefficients in the approximation \eqref{ec: x-RBM}, thus,
\begin{subequations}
    \label{ec: RBM_plasma0}
    \begin{align}
        \label{ec: RBM_plasma1}
        \mathbf{V}^\mathbf{T}\left(\mathbf{S}+\varepsilon_p(\omega_p)\mathbf{T}\right)\mathbf{V}\mathbf{\tilde{x}}(\omega_p)&=\mathbf{V}^T\mathbf{b} \\
        \label{ec: RBM_plasma}
        \left(\mathbf{\tilde{S}}+\varepsilon_p(\omega_p)\mathbf{\tilde{T}}\right)\mathbf{\tilde{x}}(\omega_p)&=\mathbf{\tilde{b}}, 
    \end{align}
\end{subequations}
with $\mathbf{\tilde{S}}\equiv \mathbf{V}^T\mathbf{S}\mathbf{V}$, $\mathbf{\tilde{T}}\equiv\mathbf{V}^T \mathbf{T}\mathbf{V}$ and $\mathbf{\tilde{b}}\equiv\mathbf{V}^T\mathbf{b}$. This new reduced problem can be solved in a much less expensive way in comparison with the FEM model, saving many computational resources ($N \ll n$). 

Let us rewrite \eqref{ec: FEM_plasma} and \eqref{ec: RBM_plasma} as follows,
\begin{subequations}
    \label{ec: RBM_plasma-2-0}
    \begin{align}
        \label{ec: RBM_plasma-2-1}
        \mathbf{A}(\omega_p)\ \mathbf{x}(\omega_p)&=\mathbf{b} \\
        \label{ec: RBM_plasma-2}
        \mathbf{\tilde{A}}(\omega_p)\ \mathbf{\tilde{x}}(\omega_p)&=\mathbf{\tilde{b}},
    \end{align}    
\end{subequations}
with $\mathbf{A}(\omega_p)\equiv \mathbf{S}+\varepsilon_p(\omega_p) \mathbf{T}$ and $\mathbf{\tilde{A}}(\omega_p)\equiv\mathbf{\tilde{S}}+\varepsilon_p(\omega_p)\mathbf{\tilde{T}}$. In order for the RBM approximation to be valid, \eqref{ec: x-RBM} and \eqref{ec: RBM_plasma-2} must hold. Then, we can state that $\mathbf{\tilde{x}}(\omega_p)$ must be the vector that makes the following residual as close to zero as possible:
\begin{equation}
    \label{ec: residual}
    \begin{aligned}
    \mathbf{R}\left(\omega_p\right)&=\mathbf{b} - \mathbf{A}(\omega_p)\mathbf{V}\mathbf{\tilde{x}}(\omega_p)=\mathbf{b}-\sum_i^N\tilde{x}_i(\omega_p)\mathbf{A}(\omega_p)\mathbf{v}_i \\
    &= \mathbf{b}-\sum_i^N\tilde{x}_i(\omega_p)[\mathbf{S}\mathbf{v}_i+\varepsilon_p(\omega_p)\mathbf{T}\mathbf{v}_i]
    \end{aligned}
\end{equation}
This is equivalent to minimize the norm of this residual. This norm can be evaluated in a straightforward way using \eqref{ec: norm_residual}. In this way, once we have calculated the inner products, $(\cdot, \cdot)$, among vectors $\mathbf{b}$, $\mathbf{S}\mathbf{v}_i$ and $\mathbf{T}\mathbf{v}_i$, evaluating the norm of this residual error is a computationally simple operation that can be carried out throughout the whole plasma frequency band of interest $\mathcal{B}$ with ease \cite{delaRubia2014Reliable}.
\begin{equation}
    \label{ec: norm_residual}
    \begin{aligned}    
    \| \mathbf{R}(\omega_p)\|^2&=(\mathbf{b},\mathbf{b})\\
    &-2\sum_i^N\operatorname{Re}\{{\tilde{x}}_i^*[(\mathbf{b},\mathbf{S}\mathbf{v}_i)+\varepsilon_p^* (\mathbf{b},\mathbf{T}\mathbf{v}_i)]\}\\
    &+\sum_i^N\sum_j^N\tilde{x}_i\tilde{x}_j^*\lbrace (\mathbf{S}\mathbf{v}_i, \mathbf{S}\mathbf{v}_j)\\
    &+\varepsilon_p^* (\mathbf{S}\mathbf{v}_i,\mathbf{T}\mathbf{v}_j)+ \varepsilon_p(\mathbf{S}\mathbf{v}_j,\mathbf{T}\mathbf{v}_i)^*\\
    &+\varepsilon_p\varepsilon_p^*(\mathbf{T}\mathbf{v}_i,\mathbf{T}\mathbf{v}_j)\rbrace,
    \end{aligned}
\end{equation}
where $*$ denotes complex conjugate. 

\textcolor{black}{Thus, the procedure for selecting suitable reduced-basis vectors $\mathbf{v}_i$ and, therefore, the reliability of this approximation, is based on a greedy algorithm \cite{delarubia2009reliable} (see Algorithm \ref{alg: greedy}) that minimizes the normalized residual error, $\|\mathbf{R}\left( \omega_p)\right)\|/\|\mathbf{b}\|$ along the plasma frequency band of interest $\mathcal{B}$. In this way, we need to solve the FEM problem only at certain $\omega_p$ to build up the reduced-basis space, i.e., those that maximize the residual norm at each loop iteration in Algorithm \ref{alg: greedy}, which can be quickly calculated by means of \eqref{ec: norm_residual}. Once we have assembled the reduced-basis projection matrix $\mathbf{V}$, we can compute the EM field solution $\mathbf{x}(\omega_p)$ in the finite element space at any plasma frequency in the band of interest $\mathcal{B}$ solving \eqref{ec: RBM_plasma-2} and using \eqref{ec: x-RBM}.}
\begin{algorithm}
\caption{Greedy algorithm}
\begin{algorithmic}[1]
\State Choose $\omega_{p}$ in $\mathcal{B}$ randomly.
\State Solve the FEM problem \eqref{ec: FEM_plasma} and obtain the solution $\mathbf{x}(\omega_{p})$.
\State Set the reduced-basis projection matrix $\mathbf{V}=\operatorname{cols}\lbrace\mathbf{x}(\omega_{p})\rbrace$ and the reduced matrices $\mathbf{\tilde{S}}$, $\mathbf{\tilde{T}}$ and $\mathbf{\tilde{b}}$.
\State Solve the reduced problem \eqref{ec: RBM_plasma-2} and obtain the reduced solution $\mathbf{\tilde{x}}(\omega_p)$, $\forall \omega_p \in \mathcal{B}$.
\State Obtain the normalized residual error $\|\mathbf{R}\left(\omega_p\right)\|/\|\mathbf{b}\|$. 
\While{$\|\mathbf{R}\left(\omega_p\right)\|/\|\mathbf{b}\|>\epsilon_{tol}$}
    \State Choose $\omega_p^{\prime}=\operatorname{argmax}_{\omega_p\in\mathcal{B}}\|\mathbf{R}\left(\omega_p\right)\|/\|\mathbf{b}\|$.
    \State Solve \eqref{ec: FEM_plasma} at $\omega_p^{\prime}$ and obtain $\mathbf{x}(\omega_p^{\prime})$.
    \If {$\mathbf{x}(\omega_p^{\prime})\not \subset \operatorname{span}\{\mathbf{V}$\}}
        \State Set $\mathbf{V}=\mathbf{V}\cup \operatorname{cols}\lbrace\mathbf{x}(\omega_p^{\prime})\rbrace$.
        \State Solve the reduced problem \eqref{ec: RBM_plasma-2}.
        \State Obtain the reduced solution $\mathbf{\tilde{x}}(\omega_p)$, $\forall \omega_p \in \mathcal{B}$.
        \State Compute $\|\mathbf{R}\left(\omega_p\right)\|/\|\mathbf{b}\|$. 
    \Else
        \State\textbf{break}
    \EndIf\EndWhile
\end{algorithmic}
\label{alg: greedy}
\end{algorithm}
\section{Practical examples and numerical results}
\label{Sec-Examples}
In order to test the RBM described above, it is applied to different practical examples of Drude-like EM scatterers, namely, a plasma sphere surrounded by both vacuum and a dielectric material, a plasma cylinder embedded by the same dielectric material, and a plasma cylinder enclosed in a crystal cover, also surrounded by vacuum. Each of the analyzed structures constitute the unit-cells (\textit{meta-atoms}) that are stacked along the $x$ and $y$ directions, forming the crystal lattice. Normal incidence and vertical polarization of the EM wave is taken into account throughout the examples. RBM performance will be compared with results obtained using CST Microwave Studio \cite{CST} and our in-house FEM software. 
Our in-house C++ code for FEM simulations uses a second-order first family of N\'ed\'elec's elements \cite{Ned80, Ing06}, on meshes generated by \texttt{Gmsh} \cite{GeuR09}.
\subsection{Plasma sphere in vacuum}
\begin{figure}[tbp]
    \centering
    \includegraphics[scale = 0.14]{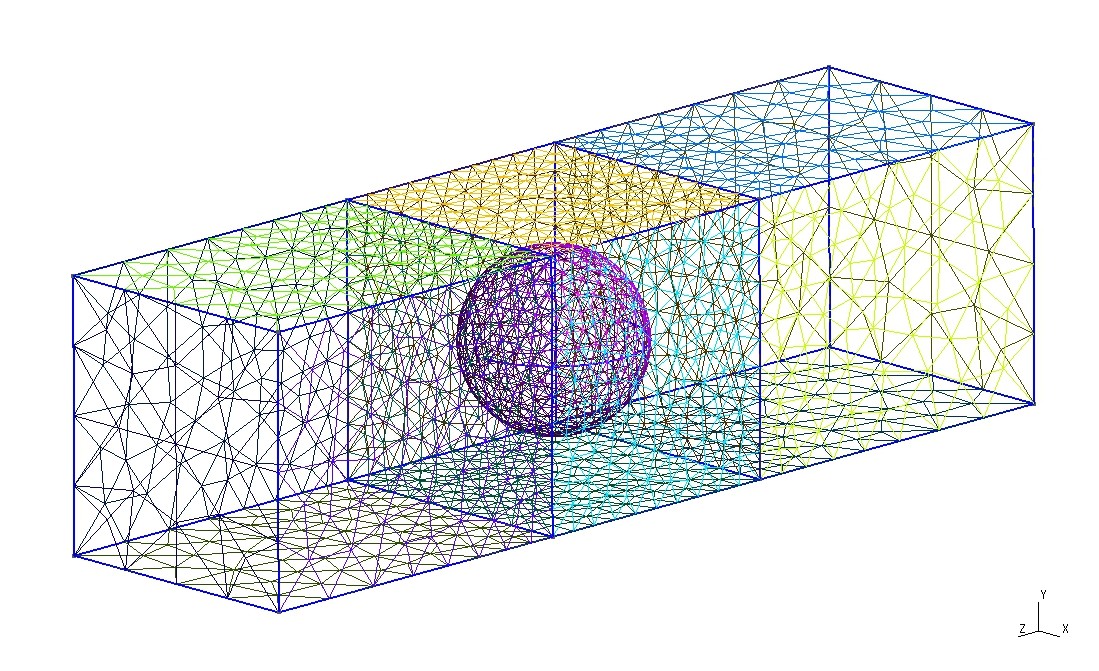}
    \caption{Plasma sphere embedded in vacuum. Sphere radius is $a = 19.965$ mm and \textcolor{black}{cell dimensions} are $L_x=L_y=L_z=3a=59.895$ mm.}
    \label{fig: sphere}
\end{figure}

\begin{figure}[tbp]
    \centering
    \includegraphics[scale = 0.7]{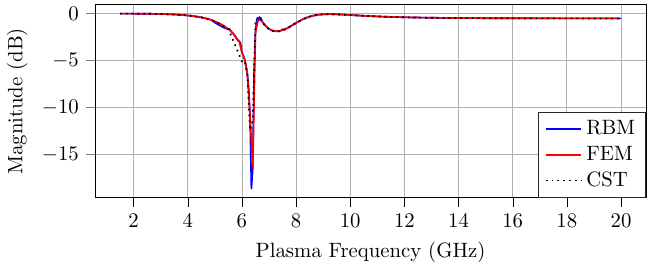}
    \includegraphics[scale = 0.7]{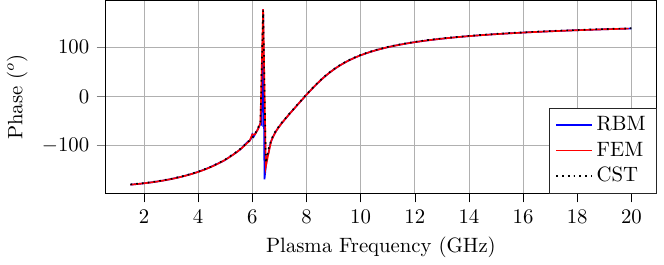}
    \caption{$S_{21}$ transfer function of the plasma sphere embedded in vacuum displayed in Fig. \ref{fig: sphere}, obtained with the \textcolor{black}{RBM presented in this work (blue), with in-house FEM (red) and with CST Microwave Studio \cite{CST} (dotted black).}}
    \label{fig: results_air_sphere}
\end{figure}

As a first example, the case of a plasma sphere embedded in vacuum is discussed. The geometry and volumetric mesh of this structure are shown in Fig. \ref{fig: sphere}. The \textcolor{black}{electromagnetic frequency} is set at $\text{f}_\text{EM}=3.8$ GHz ($\omega=2\pi \text{f}_\text{EM}$). A fast sweep of the sphere plasma frequency has been performed via RBM in the band $\mathcal{B}=[1.5, 20]$ GHz for 370 evenly spaced values. The corresponding dielectric permittivities are obtained directly from \eqref{ec: epsilon_p} with $\omega_p=2\pi \text{f}_p$ and $\gamma=10^{-3}\omega_p$. In this case, the finite element space has dimension $n=81374$. After carrying out our RBM, the obtained reduced-order model has dimension $N=16$.
In Fig. \ref{fig: results_air_sphere}, we plot the magnitude and phase of the transmission transfer function ($S_{21}$) in the plasma frequency bandwidth considered, obtained with the proposed RBM and with our in-house FEM and CST Microwave Studio. Good agreement is found.
\subsection{Plasma sphere embedded in a dielectric cube}
\begin{figure}[tbp]
    \centering
    \includegraphics[scale = 0.7]{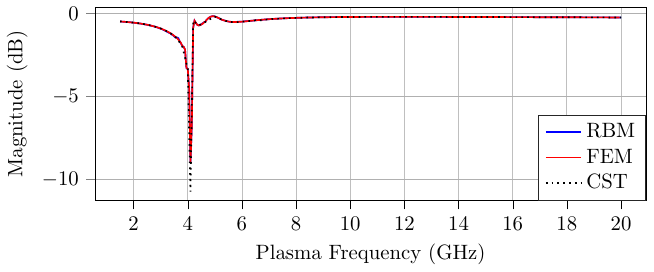}
    \includegraphics[scale = 0.7]{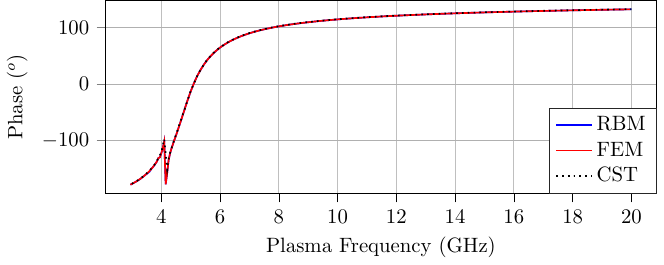}
    \caption{$S_{21}$ transfer function of the plasma sphere embedded in a dielectric material displayed in Fig. \ref{fig: sphere}, obtained with the RBM presented in this work (blue), with in-house FEM (red) and with CST Microwave Studio \cite{CST} (dotted black).}
    \label{fig: results_dielectric_sphere}
\end{figure}
As a slight modification of the previous case, we test the proposed RBM on a plasma sphere, with the same geometry as before (see Fig. \ref{fig: sphere}), but embedded in a dielectric cube material with $\varepsilon_{m}=2.96\varepsilon_0$. The electromagnetic frequency is set at $\text{f}_\text{EM}=1.8$ GHz and the plasma frequency sweep has been performed over the same band of interest as above. \textcolor{black}{Since the geometry has not changed, the FEM dimension is the same as in the previous case, $n=81374$.} The reduced-order model dimension is also $N=16$. The $S_{21}$ transfer function, obtained both via RBM and in-house FEM, along with CST Microwave Studio results, are plotted in Fig. \ref{fig: results_dielectric_sphere}. Once again, good agreement is found.

\subsection{Plasma cylinder embedded in dielectric}
\begin{figure}[tbp]
    \centering
    \includegraphics[scale=0.14]{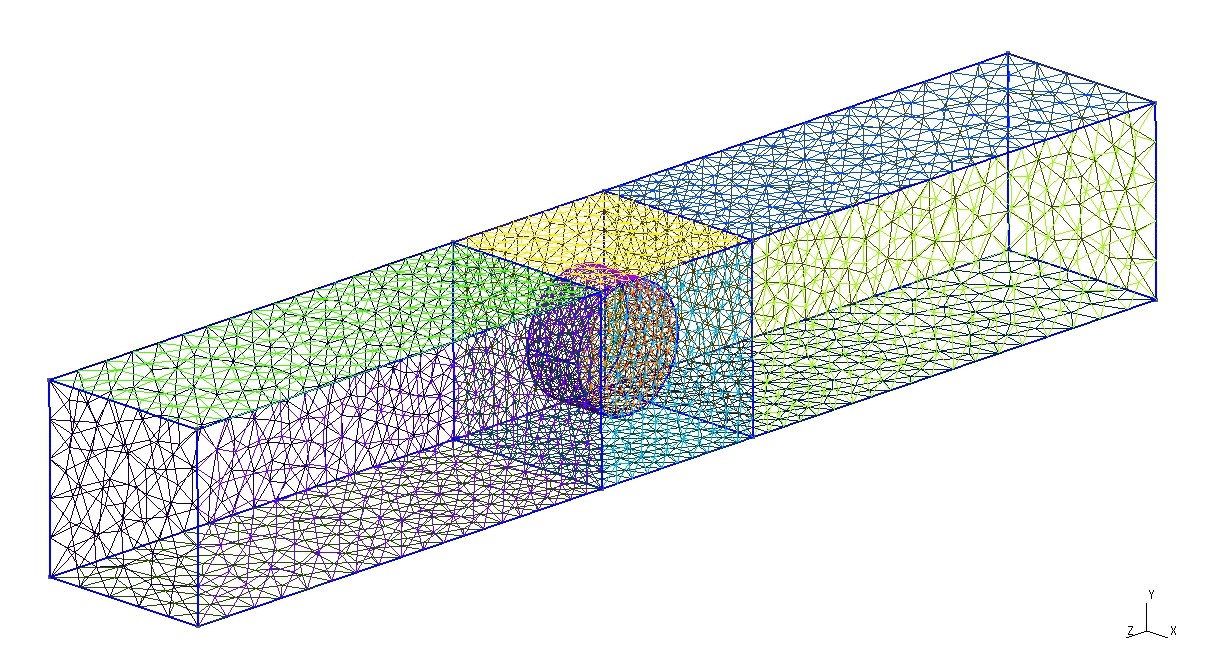}
    \caption{Plasma cylinder embedded in dielectric. Cylinder radius and height are both $a = h = 3.5$ mm and cell dimensions are $L_x=L_y=10.30$ mm and $L_z=10.50$ mm.}
    \label{fig: cylinder}
\end{figure}

\begin{figure}[tbp]
    \centering
    \includegraphics[scale = 0.7]{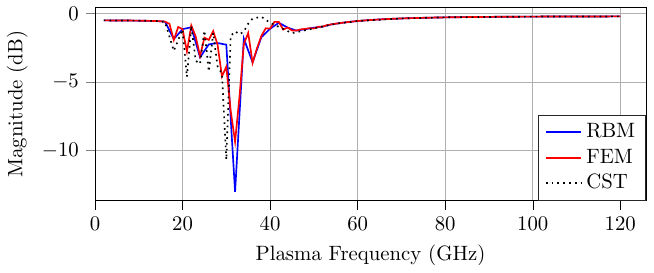}
    \includegraphics[scale = 0.7]{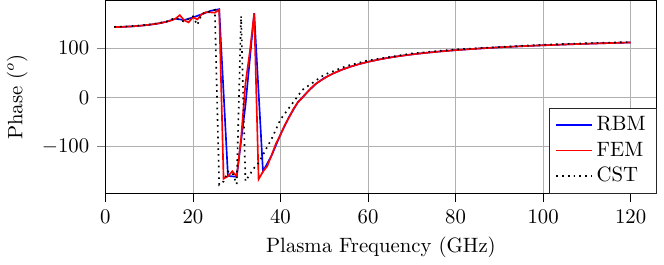}
    \caption{$S_{21}$ transfer function of the plasma cylinder displayed in Fig. \ref{fig: cylinder}, obtained with the RBM presented in this work (blue), with in-house FEM (red) and with CST Microwave Studio \cite{CST} (dotted black).}
    \label{fig: results_cylinder}
\end{figure}

In this case, we consider a plasma cylinder, depicted in Fig. \ref{fig: cylinder}, once again embedded in the same dielectric material ($\textcolor{black}{\varepsilon_{m}=2.96\varepsilon_0}$) as in the previous example. The electromagnetic frequency is now set at $\text{f}_\text{EM}=10.7$ GHz, and the plasma frequency is swept over the band $\mathcal{B}=[2,120]$ GHz with 118 evenly spaced values. The finite element problem has dimension $n = 120044$. Applying the proposed RBM, we have obtained a reduced-order model of dimension $N=19$.
The $S_{21}$ transfer function in the analyzed frequency band are plotted in Fig. \ref{fig: results_cylinder}, along with in-house FEM and CST Microwave Studio simulations. Reasonable agreement is found.
\subsection{\textcolor{black}{Plasma cylinder tubes}} 
\begin{figure}[tbp]
    \centering
    \includegraphics[scale = 0.14]{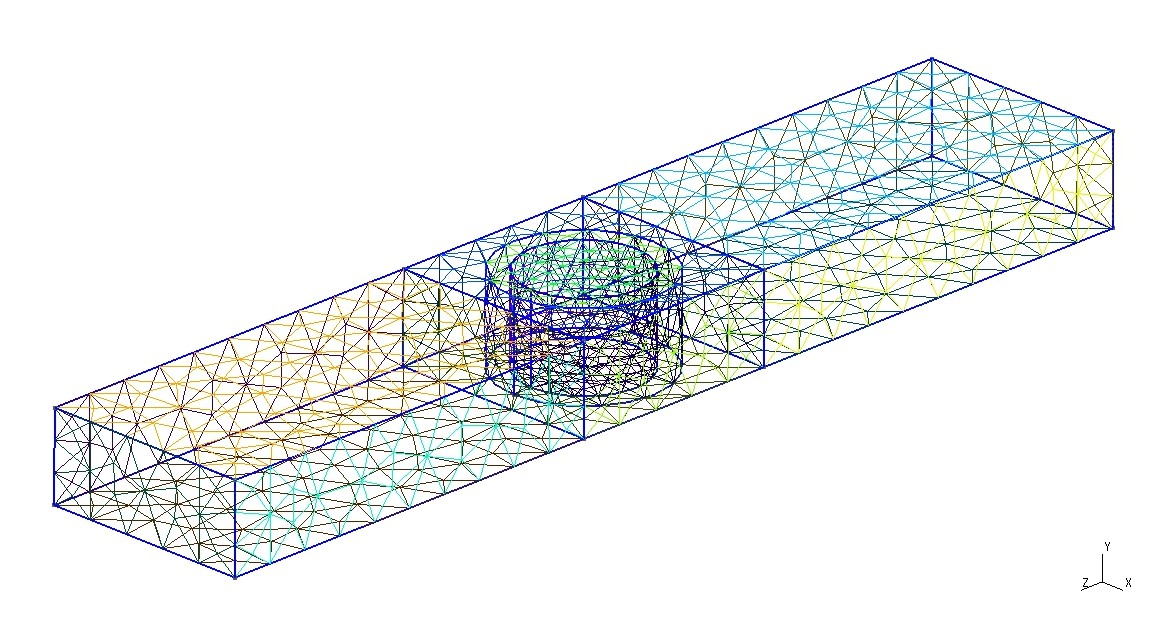}
    \caption{Plasma cylinder tube embedded in vacuum. Plasma cylinder radius and \textcolor{black}{cover internal} radius are $a=b_i=3$ mm, cover external radius is $b_e=4$ mm. They both have height $h=4.30$ mm. Initially, cell dimensions are $L_x=L_z=10.30$ mm and $L_y=4.30$ mm.}
    \label{fig: coaxial_cylinder}
\end{figure}

In this last example, we analyze a plasma cylinder surrounded by a \textcolor{black}{coaxial cylindrical cover} with $\varepsilon_m=2 \varepsilon_0$. This structure, shown in Fig. \ref{fig: coaxial_cylinder}, is embedded in vacuum, \textcolor{black}{ and it is stacked in the $x$ and $y$ directions, giving rise to a metasurface with infinitely long cylindrical tubes.} The objective is to take advantage of the fast plasma frequency sweep enabled by the proposed RBM method to efficiently and reliably obtain the EM response in various scenarios. As a result, we are able to study the dependence of magnitude and phase of the $S_{21}$ transfer function on varying metasurface features such as EM frequency or \textcolor{black}{meta-atom geometry.}

\begin{figure}[tbp]
    \centering
    \includegraphics[scale = 0.5]{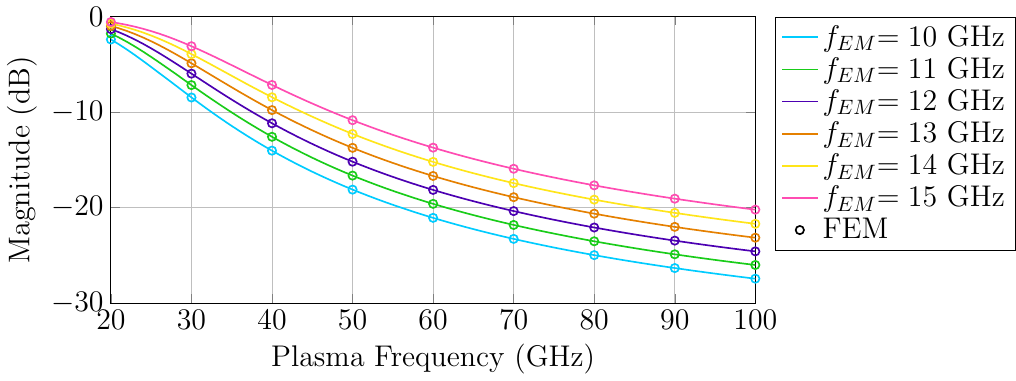}
    \includegraphics[scale = 0.5]{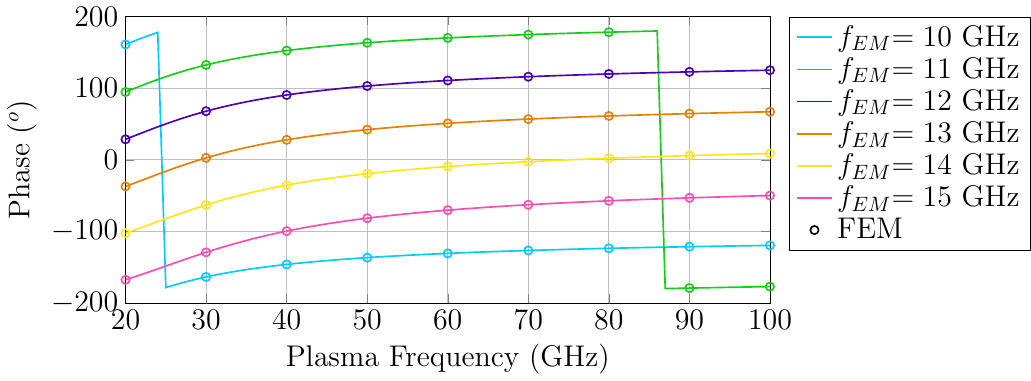}
    \caption{$S_{21}$ transfer function of the plasma cylinder tube displayed in Fig. \ref{fig: coaxial_cylinder}, obtained with the RBM presented in this work (solid), and with FEM ($\circ$) for different EM frequencies $\text{f}_\text{EM}$.}
    \label{fig: results_fEM}
\end{figure}

\begin{figure}[tbp]
    \centering
    \includegraphics[scale = 0.5]{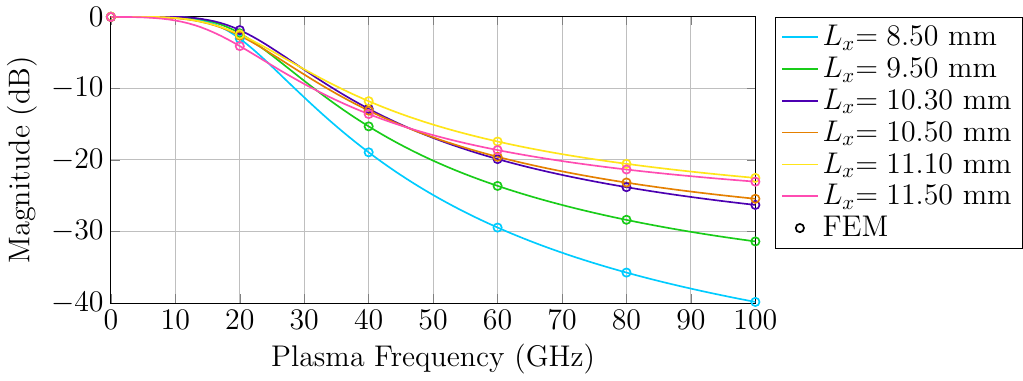}
    \includegraphics[scale = 0.5]{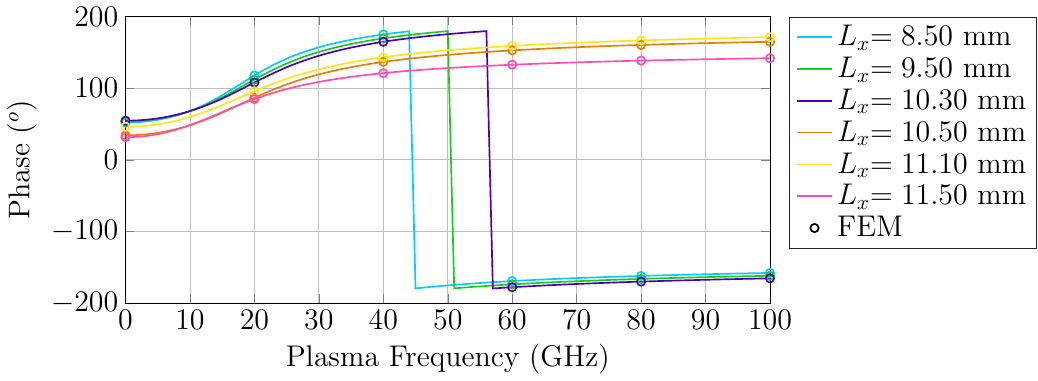}
    \caption{$S_{21}$ transfer function of the plasma cylinder tube displayed in Fig. \ref{fig: coaxial_cylinder}, obtained with the RBM presented in this work (solid), and with FEM ($\circ$) for different \textcolor{black}{length periods along the $x$ direction.}}
    \label{fig: results_xperiod}
\end{figure}

First, we perform simulations \textcolor{black}{for} different EM frequencies, $\text{f}_\text{EM}$, in the plasma frequency band $\mathcal{B}=[20,100]$ GHz with 180 evenly spaced values. $S_{21}$ results are shown in Fig. \ref{fig: results_fEM}, along with some results obtained with FEM. The FEM dimension is $n=19186$. The obtained reduced-order model has dimension $N=12,17,24$, for some values of $\text{f}_\text{EM} = 10,12,15$ GHz, respectively.
In second place, we run the simulations varying $L_x$, that is, the Drude crystal lattice period in the $x$ direction. The EM frequency is now fixed at $\text{f}_\text{EM}=10.8$ GHz. With the proposed RBM, we perform a fast plasma frequency sweep in the band $\mathcal{B}=[0,100]$ GHz with 100 evenly spaced values. The FEM dimension is $n=20784, 19186, 20058$, for some values of $L_x=8.50,10.30,11.50$ mm, respectively. RBM lowers these dimensions to $N=20,17,18$, respectively. $S_{21}$ transfer function results are detailed in Fig.~\ref{fig: results_xperiod}.
\section{Conclusions}
\label{Sec-Conclusions}
In this paper, we have presented a Reduced-Basis Method as a reliable and efficient MOR in Drude-like EM scatterers. The proposed RBM proved to be an effective approach for reducing the computational complexity of solving EM problems parametrized by the plasma frequency, while retaining a high level of accuracy. With this methodology, we are able to reduce high-dimensional FEM matrix systems (in the order of $10^5$) to lower dimensions in the order of a few tens.

By solving Maxwell's equations in this reduced-basis space, the computational cost associated with multiple plasma frequency evaluations has been significantly decreased, making the method particularly suitable for scenarios that require rapid simulations over a wide range of this parameter values. This is especially relevant in the context of adaptive design\cite{monti2024drudelikescatterers} and optimization of reconfigurable metasurfaces and metamaterials, where the plasma frequency plays a critical role in tuning the electromagnetic response.

Through various examples, we have verified the consistency of RBM with in-house FEM and commercial software, yielding very similar results. Additionally, we have successfully evaluated the EM response over a range of plasma frequencies, varying other design parameters, such as cell dimensions or electromagnetic frequency. 

The presented results have shown that RBM can achieve a good balance between computational efficiency and accuracy, being an effective tool for solving parametrized Maxwell's equations in complex scattering environments. 
\section*{Acknowledgements}
\label{Sec-Acknowledgements}
This work has been developed in the frame of the activities of the Project PULSE, funded by the European Innovation Council under the EIC Pathfinder Open 2022 program (protocol number 101099313). Project website is: \url{https://www.pulse-pathfinder.eu/}.
\bibliographystyle{sty/IEEEtran}
\bibliography{bibliography/references}
\end{document}